\documentclass[twocolumn,showpacs,preprintnumbers,amsmath,amssymb]{revtex4}

\newcommand{\simge}{ \: \raisebox{0.1cm}{\tiny $>$} \!\!\!\!
\raisebox{-0.03cm}{\tiny $\sim$} \:}

\usepackage{graphicx}
\usepackage{dcolumn}
\usepackage{bm}

\begin{document}


\title{Double-Exchange Model on Triangle Chain}

\author{Atsuo Satou}
\affiliation{
Department of Applied Physics, Tokyo University of Science,
Kagurasaka 1-3, Shinjuku-ku, Tokyo 162-8601, Japan
}

\author{Masanori Yamanaka}
 \email{yamanaka@phys.cst.nihon-u.ac.jp}
\affiliation{
Department of Physics, College of Science and Technology, 
Nihon University,
Kanda-Surugadai 1-8, Chiyoda-ku, Tokyo, 101-8308, Japan
}

\date{\today}

\begin{abstract}
We study ground state properties of the double-exchange model
on triangle chain in the classical limit on $t_{2g}$ spins.
The ground state is determined by a competition among
the kinetic energy of the $e_g$ electron,
the antiferromagnetic exchange energy between the $t_{2g}$ spins,
and frustration due to a geometric structure of the lattice.
The phase diagrams are obtained numerically for two kinds of the models
which differ only in the transfer integral being real or complex.
The properties of the states are understood
from the viewpoint of the spin-induced Peierls instability.
The results suggest the existence of a chiral glass phase
which is characterized by a local spin chirality
and a continuous degeneracy.
\end{abstract}


\maketitle

\section{\label{sec:level1}Introduction}

The double-exchange mechanism~\cite{REFzener}
has received a special attention as one of the canonical mechanisms
which explain the magnetism and transport properties
of a class of transition metals.
The Hamiltonian which describes it is known 
as the double-exchange model~\cite{REFanderson,REFdegennes}.
Assuming a two sub-lattice structure,
de Gennes obtained the phase diagram 
as a function of temperature and hole concentration,
$x$~\cite{REFdegennes}. 
The ground state exhibits the antiferromagnetic (AF),
ferromagnetic, and canted AF phases~\cite{REFdegennes,
REFnagaev}.
The disordered, two sub-lattice, and helical spin structures
and their degeneracies were discussed~\cite{REFdegennes}
combining the two sub-lattice structure.
These rich structures are induced from competition 
between the kinetic energy of the $e_g$ electron 
and the direct AF coupling between $t_{2g}$ 
spins~\cite{REFdegennes,REFgolosov,REFalonso}.

On the other hand, studies without imposing the restriction 
to the two sub-lattice structure have been reported. 
These results suggest the existence 
of the ground state without the translation invariance. 
An existence of the phase separation was 
reported~\cite{REFnagaev,REFym,REFkkm,REFgolosov2,REFkhom}.
(For the model without the direct exchange between $t_{2g}$ spins see 
refs.~\cite{REFnagaev2,REFriera,REFguinea,REFdagotto1,REFdagotto2,
REFmyd,REFsarma1,REFsarma2,REFyin}.)
The analysis based 
on the spin-induced Peierls instability~\cite{REFykm,REFKoshibae}
suggest the existence of a super-cell structure of the localized spins.
The spin state is distorted with a period which is 
commensurate with the Fermi momentum so as to open the Peierls gap,
thus stabilizing the system.
This is isomorphic to the Peierls instability~\cite{REFpeielrs}.
The only difference is that the spin degree of freedom is distorted,
instead of the lattice degree of freedom.

In two or more dimensions, the degree of the phase factor 
in the transfer integral plays an important role~\cite{REFdm,REFcb}.
For example, the generalized Peierls instability~\cite{REFhlrw} 
is expected to work and the ground state exhibits 
a quantum Hall effect~\cite{REFykm} 
if the state satisfies conditions~\cite{REFsy}.
(See also ref~\cite{REFomn}.)
Actually the staggered $\pi$-flux state is stabilized at half-filling 
in the two dimensional square lattice~\cite{REFykm,REFyunokiflux}.
The anomalous Hall effect induced from
the Berry phase was investigated~\cite{REFkmms,REFbrey3,REFomn}.
The extensive investigations 
for the phase diagram were performed in two or three dimensions
\cite{REFAnaaa,REFAlonso2,REFAlonso3}
and a flux state with super-cell structure 
was found away from half-filling.

Recently the most intriguing is the study within the framework
of Tsallis nonextensive statistics~\cite{REFreis1,REFreis2,REFreis3}.

\noindent
\begin{figure}[b]
\includegraphics[width=8cm]{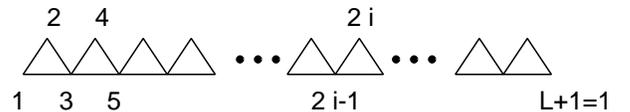}
\caption{Lattice structure of triangle chain.}
\label{fig:lattice}
\end{figure}

In this paper, we study the double-exchange model on the triangle chain
to investigate the effects of (i) the phase factor in the transfer integral
and (ii) the frustration due to the geometric structure of the lattice.
The triangle chain is constructed from the one-dimensional array 
of the triangles. (See Fig.\ref{fig:lattice}.)
The triangle is the smallest unit for making a closed loop
which is responsible for a non-vanishing flux 
up to the local gauge transformation.
Then, the triangle chain has the simplest lattice structure
to study the effect of the phase factor.
On the other hand, the geometric structure induces a frustration.
The ground state is determined by the competition among 
the kinetic energy of the $e_g$ electron, 
the antiferromagnetic exchange energy of the $t_{2g}$ spins,
and the frustration.
We numerically obtain the phase diagram
assuming the absence of the phase separation.
The ground state exhibits an insulator with a super-cell structure
of the $t_{2g}$ spins for a wide range of the parameter space.
The state has a continuous degeneracy
which is a hallmark of the double-exchange model in one dimension.
The spin structure is characterized 
by an incommensurate spin configuration \cite{REFcommentincome} 
and a local spin chirality.
The spin chirality induces a non-vanishing flux
to the $e_g$ electrons.
This is an interesting example of the finite spin chirality 
induced in electronic models without any assistance of other effects,
such as the anisotropy in spin space, spin-orbit coupling,
lattice distorsion, and external magnetic field, etc..~\cite{REFkawamura}

\section{\label{sec:level2}Hamiltonian}

The double-exchange model on the triangle chain is defined by
\begin{eqnarray}
& &H=-\Big(
\sum_{i=1}^{L}     t_{i,i+1}^{\mu}
                   c_i^{\dagger}c_{i+1}^{\phantom{\dagger}}
+ \sum_{i=1}^{L/2} t_{2i-1,2i+1}^{\mu}
                   c_{2i-1}^{\dagger}c_{2i+1}^{\phantom{\dagger}}
                   + h.c.\Big)
\nonumber\\ 
& &\hspace{10mm}+J\sum_{\langle i,j \rangle}\vec{S_i}\cdot\vec{S_j}, 
\label{eq:DE1}
\end{eqnarray}
where $c_{i,\sigma}$ is the annihilation operator 
of the $e_g$ electron at the site $i$, and
$\vec{S}_i$ are localized ($t_{2g}$) spins 
which are treated as classical vectors directed along
$(\theta_i, \phi_i)$ in the spherical coordinates.
We denote the spin configuration by a set 
$\{\theta_i, \phi_i\}$ ($i=1, 2, \cdots, L$).
Moreover, $J($ $\ge$ $0)$ is the direct exchange coupling strength 
between $t_{2g}$ spins.
$L$ (even) is the total number of the sites.
The number of the triangle is $L/2$.
In order to see the effect of the phase factor in the transfer integral,
we study two kinds of the model.
Their transfer integrals are given by~\cite{REFNagaev,REFkogan,REFmillis} 
\begin{eqnarray} 
t^{\mu=C}_{i,j}&\equiv& z_{i,j}=\cos{\frac{\theta_i}{2}}
                        \cos{\frac{\theta_j}{2}}
+e^{-i( \phi_i-\phi_j)} \sin{\frac{\theta_i}{2}}
                        \sin{\frac{\theta_j}{2}}
\label{eq:complexhoping}\\
t^{\mu=R}_{i,j}&\equiv&\vert z_{i,j} \vert=\cos{\frac{\Theta_{ij}}{2}}
\label{eq:realhoping}
\end{eqnarray} 
for C- and R-models, respectively.
Here $\Theta_{ij}$ is the relative angle 
between the spins, $\vec{S}_i$ and $\vec{S}_j$.
We impose the periodic boundary condition 
for the $e_g$ electron, $c_{L+1}=c_1$.

\begin{figure}[b]
\includegraphics[width=8cm]{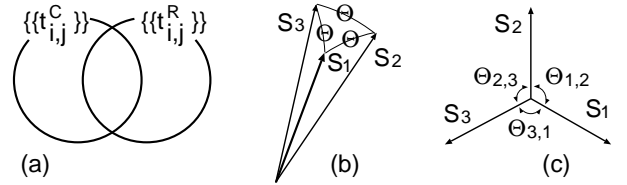}
\caption{(a) The relation between sets, $\{\{t_{i,j}^C\}\}$
and $\{\{t_{i,j}^R\}\}$.
(b), (c) Examples of $t_{2g}$ spin configuration which realize the elements
in $\{\{t^C_{ij}\}\}^c\cap\{\{t^R_{ij}\}\}$.
(b) The $e_g$ electron gets complex transfer integrals 
by a motion $1 \to 2 \to 3 \to 1$ and
(c) it gets real transfer integrals by the same motion.
The three spins are coplanar in (c).
}
\label{fig:relationzt}
\end{figure}

For a given spin configuration, 
$\{\theta_i, \phi_i\}$ ($i$ $=$1,2,$\cdots$,$L$),
the transfer integrals, $\{t^{\mu}_{ij}\}$, are uniquely determined
through (\ref{eq:complexhoping}) or (\ref{eq:realhoping}).
When we change the spin configuration, $\{\theta_i, \phi_i\}$,
the range of $\{t^{\mu}_{ij}\}$ forms a set $\{\{t^{\mu}_{ij}\}\}$.
We study the difference of the ranges between 
$\{t^C_{ij}\}$ and $\{t^R_{ij}\}$,
for the same domain,
$\{\{\theta_i, \phi_i\}\vert$
$0$ $\le$ $\theta_i$ $\le$ $\pi$,$0$ $\le$ $\phi_i$ $<$ $2\pi$,
$i$ $=$ $1$,$2$,$\cdots$,$L\}$,
or, in other words, the difference between the sets
$\{\{t^C_{ij}\}\}$ and $\{\{t^R_{ij}\}\}$.
For a system with two sites, 
the parameter space, $t^C_{1,2}=z_{1,2}$,
contains that of $t^R_{1,2}=\vert z_{1,2}\vert$,
i.e. $\{t^C_{1,2}\}$ $\supset$ $\{t^R_{1,2}\}$.
This is because
$t^C_{1,2}$ $=z_{1,2}$ $=\vert z_{1,2} \vert e^{i\Phi}$, 
$\Phi$ $=\arg[\tan(\Im z_{1,2}/\Re z_{1,2})]$,
and we can set $\Phi\equiv 0$ without loss of generality.
{\em However, for any lattice (except for tree) 
the transfer integrals (\ref{eq:complexhoping})
and (\ref{eq:realhoping}) have the mutually exclusive 
parameter ranges, i.e. $\{\{t^C_{ij}\}\}\cap\{\{t^R_{ij}\}\}^c\ne\phi$
and $\{\{t^C_{ij}\}\}^c\cap\{\{t^R_{ij}\}\}\ne\phi$},
where $A^c$ is the complement of the set $A$.
(See Fig.\ref{fig:relationzt}(a).)
The latter, $\{\{t^C_{ij}\}\}^c\cap\{\{t^R_{ij}\}\}\ne\phi$,
is not obvious.
(This holds even for the simple one dimensional chain 
with a periodic boundary condition in a finite system size.
The effect can be attributed to the boundary condition 
by a local gauge transformation 
and becomes unphysical in the thermodynamic limit.)
We show that there exist at least two examples in  
$\{\{t^C_{ij}\}\}^c\cap\{\{t^R_{ij}\}\}$
for the model on a triangle lattice.
{\em This holds for any lattice except for tree.}
The first example is 
\begin{eqnarray}
t^R_{1,2}=t^R_{2,3}=t^R_{3,1}=\cos\frac{\Theta}{2}.
\label{eq:real1}
\end{eqnarray}
We set the relative angles, $\Theta$, small and the same for simplicity.  
(See Fig.\ref{fig:relationzt}(b) for the corresponding 
$t_{2g}$ spin configuration.)
We show that $\{\{t^C_{ij}\}\}$ do not contain (\ref{eq:real1})
as a subset for any value of 
$(\theta_1, \phi_1$, $\theta_2, \phi_2$, $\theta_3, \phi_3)$.
{\em Proof}: 
We can set $(\theta_3,\phi_3)$ $=(0,0)$ without loss of generality
and get $t^C_{2,3}=t^C_{3,1}=\cos(\Theta/2)$ by setting
$(\theta_1,\phi_1$,~$\theta_2,\phi_2)$ $=(\Theta, \phi_1$,~$\Theta, \phi_2)$.
The rest parameter can be written by
$t^C_{1,2}$ $=\cos(\Theta/2) \cos(\Theta/2)$
$+ e^{-i ( \phi_1 - \phi_2)}
\sin(\Theta/2)\sin(\Theta/2)$,
which never becomes $\cos(\Theta/2)$ for $\phi_1 - \phi_2=0$.
Because, in order to make $t^C_{1,2}$ real, we have to set 
$\phi_1 - \phi_2=n\pi$ ($n=$integer) which inevitably leads to
$t^C_{1,2}=1$ or $\cos\Theta$.
This contradicts (\ref{eq:real1}).
{\large $\Box$} \ \

\noindent

One might consider that this property is due to
a finite Peierls phase along the path $1\to 2\to 3 \to 1$,
because the solid angle among 
$\vec{S}_1$, $\vec{S}_2$, and $\vec{S}_3$ is quivalent 
to the magnitude of the flux penetrating the triangle.
However, this naive observation fails.
Even when the three spins are coplanar,
there exist an element in $\{\{t^C_{ij}\}\}^c\cap\{\{t^R_{ij}\}\}$.
The second example is
\begin{eqnarray}
t^R_{1,2}=\cos\frac{\Theta_{1,2}}{2},\ \ 
t^R_{2,3}=\cos\frac{\Theta_{2,3}}{2},\ \  
t^R_{3,1}=\cos\frac{\Theta_{3,1}}{2},
\label{eq:real2}
\end{eqnarray}
with the restrictions 
$\Theta_{1,2}+\Theta_{2,3}+\Theta_{3,1}=2\pi$ and $t^R_{i,j}>0$.
(See Fig.~\ref{fig:relationzt}(c).)
We show that $\{\{t^C_{ij}\}\}$ do not contain the set (\ref{eq:real2})
as a subset for any value of 
$(\theta_1, \phi_1$, $\theta_2, \phi_2$, $\theta_3, \phi_3)$.
{\em Proof}: 
We set $(\theta_3,\phi_3)$
$=(0,0)$ without loss of generality
and get
\begin{eqnarray}
t^C_{2,3}&=&\cos\Theta_{2,3}/2=\cos\theta_2/2>0, 
\nonumber\\
t^C_{3,1}&=&\cos\Theta_{3,1}/2=\cos\theta_1/2>0, 
\nonumber\\
t^C_{1,2}&=&-\cos\Theta_{1,2}/2<0.
\label{eq:real22}
\end{eqnarray}
This means that the one of the three transfer integrals 
in $\{t^C_{i,j}\}$ should be negative.{\large $\Box$} \ 
The general proof is not easy.

\section{\label{sec:level3}Method}

To obtain the ground state 
we employ the method used in ref.~\cite{REFKoshibae}.
When we assume an occurrence 
of the $2k_F$ Peierls instability,
the unit cell for the transfer integral contains $2q$ sites
(or, in other words, $q$ triangles)
at the filling $x=p/2q$ where $p$ and $q$ are mutually prime.
The kinetic part in the Hamiltonian (\ref{eq:DE1}) is reduced to
\begin{eqnarray}
H_t&=& - \sum_{j=1}^{L/2q}
\Big( \sum_{i=1}^{2q} 
t_{i,i+1}^{\phantom{\dagger}} 
c^{\dagger}_{Nj+i-1}c_{Nj+i}^{\phantom{\dagger}} 
\nonumber\\
& &\hspace{2mm} + \sum_{i=1}^{2q/2} t_{2i-1,2i+1}^{\phantom{\dagger}} 
c^{\dagger}_{Nj+2i-1}c_{Nj+2i+1}^{\phantom{\dagger}}
+ h.c. \Big),
\label{eq:DEsymmetry}
\end{eqnarray}
where $2q$ is the number of sites contained in the unit cell
and $L$ is assumed to be the multiple of $2q$.
To obtain the ground state, we numerically optimize the functional
\begin{eqnarray}
&&E(\{\theta_i, \phi_i\})= E_t(\{\theta_i, \phi_i\})
\nonumber\\
&&+ JS^2 \sum_{\langle i,j \rangle}^{2q}
\big[\cos\theta_i\cos\theta_j 
+\sin\theta_i\sin\theta_j\cos(\phi_i-\phi_j)\big],
\label{eq:DEoptenegy}
\end{eqnarray}
where $E_t(\{\theta_i, \phi_i\})$ is 
obtained by the numerical diagonalization of (\ref{eq:DEsymmetry}).

In the calculation, the $2k_F$ spin-induced Peierls instability 
for the $e_g$ electron is assumed.
For a sufficient large $JS^2$($\simge 0.5$) and $x=0.5, 0.25$,
we confirmed that the assumption is valid in the following sense: 
We calculated the ground state energy 
for a more large super-cell structure 
which is not characterized by $2k_F$ modulation of the transfer integral
and found that the energy is the same upto 10 digits 
as that of the $2k_F$, and the spacial periodicity 
is just the repetition of that of the $2k_F$.

\section{\label{sec:level4}Phase diagram for C-model}

We show the phase diagram for C-model in Fig.\ref{fig:phasediagram}(a). 
The phases are classified by the spin configuration,
the modulation of the transfer integrals, 
and the position of the Fermi point.

\subsection{\label{sec:levelC1}Several limits}

(1) For $J=0$ the spin configuration is expected to be ferromagnetic,
because the double-exchange mechanism widens the energy band width 
of $e_g$ electrons to gain the kinetic energy. 
The ground state is metallic.

\noindent
(2) For $J=\infty$ or $x=0,1$, the spin configuration exhibits 
mutually 120-degree structure because of the frustration.

\noindent
(3) For $x=1/2$, in this model the Wigner-Sites cell contains two sites
and the state is a band insulator.

\subsection{\label{sec:levelC2}Region $JS^2\sim 0$}

We have two phases.

\noindent  
(1) For $x$ $<$ $0.5$, the state is the ferromagnetic metal (FM).
The spin configuration is ferromagnetic
and the amplitude of the transfer integrals becomes uniform.
The unit cell of the transfer integral coincides with
that of the Wigner-Sites cell.
Then we have an empty upper band and a partially filled lower band.

\noindent
(2) For $x$ $>$ $0.5$, the state exhibits an
incommensurate~\cite{REFcommentincome} band insulator (BI-1).
For $x=p/2q$, a ferromagnetic $2q$-merization occurs.
The spin configuration is completely ferromagnetic 
within the domain and incommensurate in long distance.
All the energy bands are dispersive.
The energy gap opens at the Fermi point.

\subsection{\label{sec:levelC3}Region $JS^2\to \infty$}

The spin configuration is nearly 120 degree structure
and three spins in a triangle are coplanar.

\noindent
(1) For $x$ $<$ $0.5$, the state exhibits the incommensurate
band insulator (BI-2).
All the sub-bands are dispersive.  
(This is in contrast to the simple one dimensional chain
where all the sub-bands are dispersionless~\cite{REFKoshibae}.) 

\noindent
(2) For $x$ $>$ $0.5$, the state is in the metallic phase (120M-1).
The amplitude of the transfer integral is uniform 
as a unit of the Wigner-Sites cell
and we have two dispersive energy bands.

\noindent
\begin{figure}[b]
\includegraphics[width=8cm]{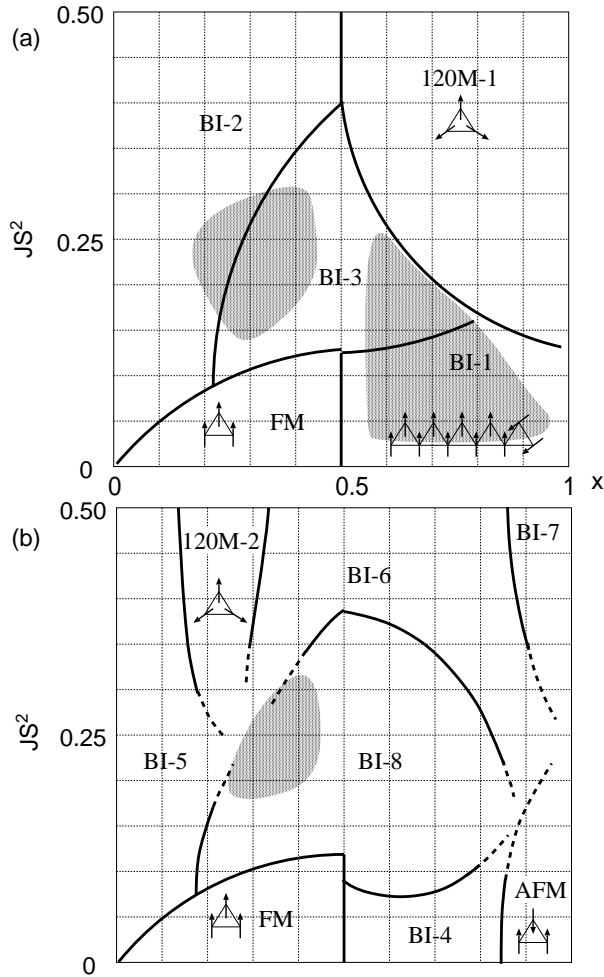}
\caption{Possible phase diagrams for C-model (a) and R-model (b)
as a function of the electron concentration $x$
and antiferromagnetic direct exchange coupling $JS^2$
between $t_{2g}$ spins.
FM is the ferromagnetic metal phase.
BI's are the incommensurate band insulators
with incommensurate ferromagnetic polymerization (BI-1),
with incommensurate spin (BI-2,4,5,6),
and with dispersionless energy band (BI-3,7).
120M-1 and 2 are the metallic phases
with nearly 120 degree structure.
AFM is the sub-lattice antiferromagnetic phase.
The masked regions are the chiral glass phases.
The calculation is performed at points where broken lines cross.
}
\label{fig:phasediagram}
\end{figure}

\subsection{\label{sec:levelC4}Region $JS^2 \sim 0.25$}

The spin configuration has a rich structure 
because of the competition between the antiferromagnetic exchange coupling
and the kinetic energy of the $e_g$ electrons 
which favors the ferromagnetic spin configuration.
The incommensurate insulator
with dispersionless energy bands (BI-3) occupies 
a large region in the center of the diagram. 
Except for $x=1/2$, the spin configuration and the transfer integral
are incommensurate.
(The property is the same that of the incommensurate gapful phase
in the simple one dimensional chain~\cite{REFKoshibae}.) 
In the masked region in the diagram,
the three spins in each triangle are not coplanar.
This phase is characterized by a finite spin chirality
and the non-vanishing flux when $e_g$ electron moves around 
one of the triangles.
(The chirality is defined by
$\sum_j\langle\vec{S}_1 \cdot (\vec{S}_2 \times \vec{S}_3)\rangle_j$,
where $j$ is the index of the triangle.)
The state exhibits continuous infinite degeneracy.
Therefore, the state is a chiral glass.
Each degenerate ground state has a finite chirality.
Because of the lattice structure,
where the adjacent two triangles share only one site,
the degenerate ground states with different magnitude of the chirality
are continuously connected to each other.
The continuous symmetry breaking is forbidden in one dimension.
Therefore, in general, the chirality vanishes 
because the ground state is a superposition of the states 
with different chiralities.

\section{\label{sec:level5}Phase diagram for R-model}

The phase diagram for R-model is shown in Fig.\ref{fig:phasediagram}(b).

\subsection{\label{sec:levelR1}Region $JS^2\sim 0$}

We have three phases.

\noindent
(1) ferromagnetic metal phase (FM) for $x$ $<$ $0.5$,

\noindent
(2) incommensurate insulating phase (BI-4) for $0.5$ $<$ $x$ $<$ $0.85$,
and

\noindent
(3) metallic antiferromagnetic phase (AFM) for $0.85$ $<$ $x$.

In BI-4 phase three spins in each triangle are coplanar
within a numerical error.
This property is different from that of BI-1 phase in C-model.
The AFM phase is sub-lattice AF 
and the transfer integrals between the sub-lattices terminate.
Then we have one cosine band
and a decoupled dispersionless energy band.

\subsection{\label{sec:levelR2}Region $JS^2\to\infty$}

The spin configuration is nearly 120 degree structure
and three spins in each triangle are coplanar.
There are three phases:

\noindent 
(1) metallic phase (120M-2) for $0.15$ $<$ $x$ $<$ $0.35$,

\noindent
(2) incommensurate insulator for $0$ $<$ $x$ $<$ $0.15$ (BI-5) 
and $0.35$ $<$ $x$ $<$ $0.85$ (BI-6), 
and 

\noindent
(3) incommensurate insulator with dispersionless energy band (BI-7)
for $0.85$ $<$ $x$.

The difference between 120M-1 and 120M-2 phases is as following:
The transfer integrals are able to be negative
in 120M-1 whereas all the transfer integrals in 120M-2 are positive.
The case holds between BI-5, 6, and BI-2 phases.
We could not find the difference between BI-5 and 6 phases.
In BI-7 the spin configuration and the transfer integral 
are incommensurate.
In this phase all energy bands are dispersionless 
because some transfer integrals vanish due to the spin configuration

\subsection{\label{sec:levelR3}Region $JS^2 \sim 0.25$}

The incommensurate insulator
with dispersionless energy band (BI-8)
occupies large region in the center of the diagram except for $x=1/2$.
The transfer integral exhibits a modulation.
The masked region is the chiral glass phase.
(In the R-model, the transfer integral is replaced by it's absolute value
so the $e_g$ electron do not be affected by the phase factor.)
In the vicinity of full filling,
there are several phase boundaries of five phases,
BI-4,6,7,8 and AFM.
In low dope region, we have four phases,
BI-5,6,8 and 120M-2.

\noindent
\begin{figure}[t]
\includegraphics[width=8cm]{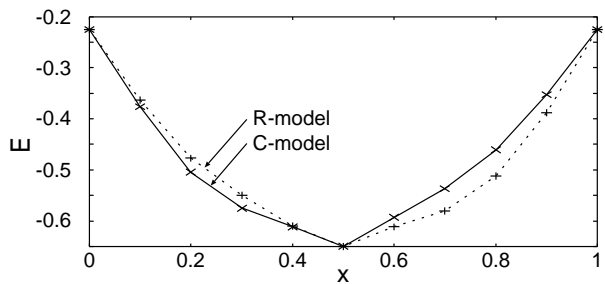}
\caption{Estimates of the ground state energy for 
C-model (solid line) and R-model (broken line)
as a function of electron concentration $x$ at $JS^2=0.3$.
The error bars are smaller than the plotted points.
}
\label{fig:energystruc}
\end{figure}

\section{\label{sec:level6}Discussion}

We have obtained a possible phase diagram of the double-exchange model
on the triangle chain at zero temperature.
The ground state has a rich structure,
which includes a band insulator 
with an incommensurate spin structure~\cite{REFcommentincome}
and a chiral glass phase.
These states exhibit the continuous degeneracy.
One of the origins of the degeneracy is due to the structure of the lattice
where the adjacent triangles share only one site.
The phase diagram is different from the simple one dimensional chain
mainly in the following two points:
(i) The structure of the phase diagrams 
of C- and R-models is different.
(In the simple one dimensional chain, it is the same.)
This is due to the enclosure relation 
between $\{\{t^C_{ij}\}\}$ and $\{\{t^R_{ij}\}\}$.
The transfer integral of BI-2 belongs
to $\{\{t^C_{ij}\}\}\cap\{\{t^R_{ij}\}\}^c$,
those of BI-7 and BI-6 ($x$ $>$ $0.5$) belong
to $\{\{t^C_{ij}\}\}^c\cap\{\{t^R_{ij}\}\}$,
and those of FM and the states at $x$ $=$ $0.5$ 
belong to $\{\{t^C_{ij}\}\}\cap\{\{t^R_{ij}\}\}$.
Reflecting this property, the different ground states are stabilized
in R- and C-models in the same parameter in $(x, J)$ space. 
For a sufficient large $JS^2$,
it is found that the ground state energy of C-model 
is lower than that of R-model for $x$ $<$ $0.5$ 
and vice versa for $x$ $>$ $0.5$.
The example is shown in Fig.\ref{fig:energystruc}.
(Within our analysis, the energy dependence as a function 
of the electron concentration is concave near $x\sim 0$ and $1$.
This suggests the absence of the phase separation.
However, we need further investigation for the definite result.)
Then we may be careful when we substitute or approximate
(\ref{eq:complexhoping}) by (\ref{eq:realhoping}).
(ii) For a small $J$, the global structure of the phase diagram
is similar to that of the simple one dimensional chain. 
However, when $J$ is large enough, the 120-degree structure 
is dominant because of the frustration.
Thus the AF phase which is appeared 
in the simple one dimensional chain vanishes.

We use the numerical optimization method to obtain
the ground state energy.
In general, the method is known to be difficult 
to the function with many variables.
We have two sites in the unit cell 
and two parameters, $\{\theta, \phi\}$, for each site.
Therefore, we have $4q$ independent parameters
which should be optimized for an electron concentration
$x=p/q$. 
The inverse of the denominator is the resolution
of the electron concentration in the phase diagram. 
Actually, for $x=1/10$ we have 40 parameters
and the number of parameter is a realistic limit 
of our numerical calculation.
We cannot examine the electron concentration 
with more large denominator by the present method.
The connectivity of the critical lines and the precise position
of the phase boundaries are not clear 
due to the above difficulty.
The simple extension of the method
for many parameters seems not to be realistic.

\begin{acknowledgments}
The computation in this work has been done using the facilities 
of the Supercomputer Center, Institute for Solid State Physics, 
University of Tokyo 
and Yukawa Institute for Theoretical Physics.
This work is supported by Research Grant for Assistants 
(and Young Researchers) of Nihon University Research Grant 
for 2002.
\end{acknowledgments}

\newpage


\begin{thebibliography}{99}
\bibitem{REFzener}
C.~Zener,  
Phys. Rev. {\bf 82}, 403 (1951). 

\bibitem{REFanderson}
P.W.~Anderson and H.~Hasegawa:
Phys. Rev. {\bf 100}, 675 (1955).

\bibitem{REFdegennes}
P.-G.~de Gennes, Phys. Rev. {\bf 118}, 141 (1960).

\bibitem{REFnagaev}
\'E.L.~Nagaev, Soviet Phys. JETP, {\bf 30}, 693 (1970).

\bibitem{REFgolosov}
D.I.~Golosov, M.R.Norman, and K.~Levin,
Phys. Rev. {\bf B58} 8617 (1998).

\bibitem{REFalonso}
J.L.~Alonso, L.A.~Fern\'andez, F.~Guinea, V.~Laliena,
and V.~Mart\'in-Mayor, 
Phys. Rev. {\bf B63} 054411 (2001).

\bibitem{REFNagaev}
\`E.L.~Nagaev, {\em Physics of Magnetic Semiconductors}
Moscow, Mir Publ.,1979.

\bibitem{REFNagaev5}
\`E.L.~Nagaev, Sov. Phys.-Uspekhi {\bf 166} 833 (1996).

\bibitem{REFym}
S.Yunoki and A.Moreo,
Phys. Rev. {\bf B58} 6403 (1998).

\bibitem{REFkkm}
M.Yu.~Kagan, D.I.~Khomskii, and M.~Mostovoy,
Eur. Phys. J. {\bf B12} 217 (1999).

\bibitem{REFkhom}
D.I.~Khomskii, cond-mat/9909349.

\bibitem{REFgolosov2}
D.I.~Golosov,
J. Appl. Phys. {\bf 91} 7508 (2002).

\bibitem{REFnagaev2}
\`E.L.~Nagaev, Physica B{\bf230-232}, 816 (1997).

\bibitem{REFriera}
J.~Riera, K.~Hallberg and E.~Dagotto, 
Phys. Rev. Lett.{\bf 79}, 713 (1997).

\bibitem{REFguinea}
D.P.~Arovas and F.~Guinea 
Phys. Rev. {\bf B58}, 9150 (1998).

\bibitem{REFdagotto1}
S.~Yunoki, J.~Hu, A.L.~Malvezzi, A.~Moreo, N.~Furukawa, and E.~Dagotto,
Phys. Rev. Lett.{\bf 80}, 845 (1998).

\bibitem{REFdagotto2}
E.~Dagotto, S.~Yunoki, A.L.~Malvezzi, A.~Moreo, J.~Hu, S.~Capponi, D.~Poilblanc
and N.~Furukawa, Phys. Rev. {\bf B58}, 6414 (1998).

\bibitem{REFmyd}
A.~Moreo, S.~Yunoki, and E.~Dagotto,
Science {\bf 283}, 2034 (1999).

\bibitem{REFsarma1}
A.~Chattopadhyay, A.J.~Millis, and S.Das Sarma,
Phys. Rev. B{\bf 61}, 10738 (2000).

\bibitem{REFsarma2}
A.~Chattopadhyay, A.J.~Millis, and S.Das Sarma,
Phys. Rev. B{\bf 64}, 012416 (2001).

\bibitem{REFyin}
L.~Yin,
Phys. Rev. B{\bf 68}, 104433 (2003).

\bibitem{REFykm}
M.~Yamanaka, W.~Koshibae, and S.~Maekawa,
Phys. Rev. Lett. {\bf 81}, 5604 (1998).

\bibitem{REFKoshibae}
W.~Koshibae, M.~Yamanaka, M.~Oshikawa, and S.~Maekawa,
Phys. Rev. Lett. {\bf 82}, 2119 (1999).

\bibitem{REFpeielrs}
R.E.~Peierls, {\it Quantum Theory of Solids},
Clarendon Press, Oxford, 1955.

\bibitem{REFdm}
E.~M\"uller-Hartmann and E.~Dagotto, 
Phys. Rev. B{\bf 54}, R6819 (1996).

\bibitem{REFcb}
M.J.~Calder\`on and L.~Brey,
Phys. Rev. B{\bf 58}, 3286 (1998).

\bibitem{REFhlrw}
Y.~Hasegawa, P.~Lederer, T.M.~Rice, and P.B.~Wiegmann,
Phys. Rev. Lett. {\bf 63}, 907 (1989).

\bibitem{REFsy}
A.~Satou and M.~Yamanaka,
cond-mat/0001314.

\bibitem{REFomn}
K.~Ohgushi, S.~Murakami, and N.~Nagaosa,
Phys. Rev. B{\bf 62}, R6065 (2000).

\bibitem{REFyunokiflux}
D.F.~Agterberg and S.~Yunoki,
Phys. Rev. B{\bf 62}, 13816 (2000).

\bibitem{REFkmms}
Y.B.~Kim, P.~Majumdar, A.J.~Millis, and B.I.~Shraiman,
cond-mat/9803350;
J.~Ye, Y.B.~Kim, A.J.~Millis, B.I.~Shraiman, 
P.~Majumdar, and Z.~Tesanovic,
Phys. Rev. Lett. {\bf 83}, 3737 (1999).

\bibitem{REFbrey3}
M.J.~Calderon and L.~Brey,
Phys. Rev. B{\bf 63}, 054421 (2001).

\bibitem{REFAnaaa}
H.~Aliaga, B.~Normand, K.~Hallberg, M.~Avignon, and B.~Alascio
Phys. Rev. B{\bf 64}, 024422 (2001).

\bibitem{REFAlonso2}
J.L.~Alonso, L.A.~Fern\`andez, F.~Guinea, V.~Laliena, 
and V.~Mart\`in-Mayor
Phys. Rev. B{\bf 63}, 064416 (2001).

\bibitem{REFAlonso3}
J.L.~Alonso, J.A.~Capitan, L.A.~Fern\`andez, F.~Guinea, and
V.~Mart\`in-Mayor
Phys. Rev. B{\bf 64}, 054408 (2001).

\bibitem{REFreis1}
M.S.~Reis, J.C.C.~Freitas, M.T.D.~Orlando, E.K.~Lenzi,
and I.S.~Oliveira, Europhys. Lett. {\bf 58}, 42 (2002).

\bibitem{REFreis2}
M.S.~Reis, V.S.~Amaral, J.P.~Ara\`ujo, and I. S.~Oliveira,
Phys. Rev. B{\bf 68}, 014404 (2003).

\bibitem{REFreis3}
M.S.~Reis, J.P.~Araujo, V.S.~Amaral, E.K.~ Lenzi, 
and I.S.~Oliveira,
Phys. Rev. B{\bf 66}, 134417 (2002).

\bibitem{REFcommentincome}
The ``incommensurate'' means
that the spin configuration is characterized 
by a $k_F$ incommensurate configuration.
The corresponding transfer integral has a $2k_F$ modulation.

\bibitem{REFkawamura}
For a review, see  
H.~Kawamura, J. Phys.: Condens. Matter {\bf 10} 4707 (1998)
and references therein.

\bibitem{REFkogan}
E.M.~Kogan and M.I.~Auslender,
Physica Status Solidi B{\bf 147}, 613 (1988).

\bibitem{REFmillis}
A.J.~Millis, P.B.~Littlewood, and B.I.~Shraiman,
Phys. Rev. Lett. {\bf 74}, 5144 (1995).


\end{thebibliography}
\end{document}